# Ambipolar perovskite light electrochemical cell for transparent display devices

Arthur Ishteev*; Ross Haroldson[1]; Dmitry Gets[2]; Alexey P. Tsapenko[3]; Masoud Alahbakhshi[4]; Jiyoung Moon[5]; Patricia Martinez[6]; Khabib Usupov[7]; Albert Nasibulin[3]; Sergey Makarov[2]; Anvar Zakhidov[6]

*Corresponding author arthurishteev@misis.ru
Laboratory of Advanced Solar Energy, NUST MISiS, Moscow, Russia
[1] Department of Physics, The University of Texas at Dallas, Richardson, Texas, United States
[2] Department of Nanophotonics and Metamaterials, ITMO University, St. Petersburg, Moscow 197101, Russia
[3] Skolkovo Institute of Science and Technology, Nobel str. 3, Moscow, Russian Federation
[4] Department of Electrical and Computer Engineering, The University of Texas at Dallas, Richardson, Texas, United States
[5] Department of Materials Science and Engineering, The University of Texas at Dallas, Richardson, Texas, United States
[6] NanoTech Institute, The University of Texas at Dallas, Richardson, Texas, United States
[7] Department of Engineering Sciences and Mathematics Luleå University of Technology, Luleå, Sweden

## Introduction

Perovskites — a broad category of compounds that share a certain crystal structure $ABX_3$. Lead-halide perovskites are low-cost, solution-processable materials with excellent intrinsic properties such as broad tunability of bandgap across the visible spectrum, low exciton-binding energy (9–80 meV), low trap-state density ($10^{15}$–$10^{17}$ cm$^{-3}$), high photoluminescence quantum efficiency [1] and high emission color purity (narrow full-width at half maximum). The remarkable progress of perovskite solar cell [2] [3] was initially driven by rapid empirical improvements and shown versatile functionality for a variety optoelectronic devices, such as perovskite light-emitting diodes [4] or light electrochemical cells (PeLEC). Initial rationales for efficiency improvements included morphological control and charge confinement through halide perovskite nanocrystals [5] led to enhancement of charge injection and balance [6] [7].

Perovskite light-emitting devices exceeded 20% quantum efficiency and low turn-on voltage 1.25V [8], which is highly beneficial for the high-definition display technology. Low costs owing to solution fabrication process and simplicity of the device architecture have been actively investigated for potential commercialization [9] as display devices.

Global perspectives on consumable electronic market demonstrate trends to expansion of high resolution displays for wearable devices [10] [11]. Currently the majority commercial displays have side-by-side red, green and blue (RGB) subpixel layout. This configuration constrains the numerical increase of pixel density, which requires to shrink of each subpixel area, which is limitation to high resolution displays for virtual and augmented reality (AR). Alternative layout of transparent subpixels on top of each other has a potential for new generation of long-term display devices such as AR and heads up display (HUD). Stack configuration of subpixels requires transparent and stable electrode material.

Among all outstanding properties, organohalide perovskites has an equivocal feature that is ionic migration[12][13]. Ionic migration[14] is vacancy assisted migration of halide ions. Ion redistribution have a facile activation energy (~0.3 eV) and high ionic conductivity to move within perovskite devices [14] [15]. Ion migration can be triggered either by light illumination [16] or by application of electric voltage [17] and effects on optical properties and device performance. Generally, in perovskites ionic migration results in hysteresis behavior of current-voltage characteristics [18] [12]. But ionic migration can demonstrate advantageous properties. Ion redistribution balances carrier injection, resulting in high electroluminescence efficiency [19].

Density functional theory (DFT) calculation of $CsPbBr_3$ band structure demonstrates that perovskite transitional interstitial defects can dope the host perovskite. The excess of these defects can impart p- or n-type conductivity to the perovskite [20]. Cesium ion and bromine vacancy produce donor level near the valence band at the same time Br ion and Cs vacancy produce acceptor level near the conduction band. Therefore, migration of these ions towards perovskite interfaces leads doping of these interfaces and formation of p-i-n structure inside the perovskite layer [23]. This mechanism is similar to the light-emitting electrochemical cell (LEC) and was demonstrated on perovskite monocrystal [24]. Voltage induced ionic migration has reversible behavior [14] and depends on external field. The removal of external excitation leads to ions relaxation and return the device to initial state. Voltage and light induced Ion diffusion cause reactions between perovskite and metal electrode [25] and transport layer [26] and directly effect on life time of the device.

The replacement of conventional metal electrodes to graphene related materials leads to suppression of degradation on perovskite/electrode interface and extend device lifetime [27]. Printable perovskite device with carbon electrode are ready for industrial fabrication approach [28]. Graphene interface engineering approach allows to extend operation life time [29]. Single wall carbon nanotubes (SWCNT) [30] provide hole transport properties [31], highly transparency at visible range and enhance the stability of perovskite optoelectronic devices [32], [33]. The combination of inorganic perovskite with carbon nanotubes provide an opportunity to fabricate incredibly stable, transport layer free flexible light emitting devices [29] [34].

One of the efficient methods to tune the intrinsic properties of carbon nanotubes is doping [35] Alkali metals like Cs, and halides (I, Br), can serve as n- and p-doping materials for CNTs, respectively [36], [37], [38]. The effect of doping is manifested by changing the Fermi level of carbon nanotubes as well as the work function, carrier concentration and conductivity [39]. Switching functionality requires reversibility of the doping process. In other words, one needs a way to remove dopants easily from CNTs to restore their initial properties. Such reversible doping can be implemented by controlling an external electric field, where the switchable regime is realized by inverting the direction of the field.

Here we present easy-to-do transparent ambipolar light emitting subpixel based on inorganic perovskite and single wall carbon nanotubes. This structure allows for the formation of a p-i-n junction in pero-polymer layer under applied voltage bias. The inorganic perovskite light emitting device with transparent tunable SWCNT electrode demonstrates light emission at forward and reversed bias. Voltage induced p-i-n formation in perovskite allows to manipulate device structure and polarize the device. In-situ doping of SWCNT on perovskite interface work function due to ion migration in perovskite in electrical field. The device is transparent in the visible region, that allows to assemble stack pixel of multicolor cells one above the other (figure S3). This makes smaller pixel size possible, compared to conventional side-by-side subpixel configurations.

Small bright perovskite tandem pixels provide higher display resolutions and have potential applications as micro-LED displays for augmented reality (AR) technology and head-up display (HUD). Tandem devices share a carbon electrode, which injects holes and electrons to both top and bottom subpixels.

## Discussions

The easy-to do light emitting device based on thin film of inorganic $CsPbBr_3$ and polyethylene oxide (PEO) deposited on top of glass-ITO substrate with press-transferred top single wall carbon nanotubes (SWCNT) top electrode. Perovskite layer deposition is described in the experiment section below.

Hereinafter we presume the following designation. The device was connected to source-meter unit and the following operations were applied. **Forward sweep** is periodically varying voltage, applied from 0 V to 4V which means **ITO injects holes, SWCNT injects electrons**. This regime is plotted in the first (top right) quadrant of the IV plot (positive current and voltage). **Reverse sweep** presumes that voltage applied from 0 V to -4V which means **hole injection from SWCNT and electrons from ITO**. This mode is shown in the third quadrant of the IV plot. **Polling** is constant voltage exposure for 120 seconds. We observed polling mechanism in our previous article.

The initial state for the perovskite LEC demonstrates visible electroluminescence (EL) under forward sweep from 2.4 V (Figure 1 line 1). In response to an applied bias, cations drift toward and accumulate near the cathode, and anions likewise move toward and pack near the anode. Reverse sweep does not allow go through the device due to potential barrier (Figure 1 line 2). Further consequent forward sweeps demonstrate an upward current shift related to higher injection while threshold voltage decreases until saturation. This state corresponds to the accumulation of positively charged ions and defects on perovskite-SWCNT interface (Figure 1 top).

First reverse sweeps (third quadrant: ITO injects electrons, SWCNT injects holes) demonstrate extremely weak current with no signs of EL (line 2). The device works as rectifying diode and blocks current at reverse bias. High work function prohibits hole injection from SWCNT. Multiple consequent sweeps from 0 to -4.0 V or poling at -2.3 V corresponds to highly ohmic contact and shows tendency to current shift. We assume that voltage induced ion exchange on SWCNT interface facilitates a decrease in the work function and allow reverse bias. Thus, IV curves are intermittent, and it bounces between two distinguishable states (figure S1 supplement). The bouncing is attributed to the shot noise and occurs due to unstable p-i-n junction that forms under electrical field during reverse voltage sweeps. Voltage bias over threshold voltage (-2.3V) correspond to voltage induced ion redistribution in polymer-perovskite layer and, at the same time, a work function tuning by doping of SWCNT (Figure 1) in external electrical field. When polling was applied, we observed that the device demonstrates higher current and visible EL (line 3) as bright as we observed in the beginning. In contrast, forward sweep demonstrated weak current in the first current (line 4), EL was not observed. The device was fully polarized from what we start. Full polarization cycle was completed by return the device to the initial state by the polling at +2,3V (line 5).

We purpose the combination of two mechanism of in-situ device polarizing: SWCNT work function tuning and voltage induced p-i-n/n-i-p formation.

P-i-n formation was confirmed by the electroluminescence observed for the symmetrical device structure (figure S2 in supplementary). The emissive layer was formed by two $CsPbBr_3$:PEO films over ITO electrodes facing each other on glass and polyethylene terephthalate (PET) substrates. At a point of contact between two substrates we observed green emission at +4.0V and -4.0V, that confirms voltage induced formation of p-i-n structure. In-situ formation of p-i-n structure in electrical field allows to polarize the device from p-i-n to n-i-p. Additionally we assembled the configuration equal to perovskite subcells are connected back-to-back in-series through EGaIn. This configuration is equal to two Shottky diodes based on perovskite-ITO subcells. The details are described in supplementary information (figure S3).

The CNT doping process can be divided into two stages (Figure 1). In the first stage, the negatively-charged bromide ions (anions) move towards the ITO side and the positive cesium ions (cations) aggregate at the perovskite/CNT interface when forward bias is applied. Such ionic migration inside the perovskite layer under electric poling by external field results in the formation of p-i-n homojunction structure. In the second stage, above a certain threshold, cesium cations accumulated at the area near SWCNT during positive poling start to penetrate through the interface and fill the space between SWCNT bundles or inside the carbon structure since ionic radius of cesium is sufficiently small. Such adsorption of positive ions is n-type doping of SWCNT that changes the electrical properties of electrode: the work function of SWCNTs is shifted downward. This process facilitates the injection of electrons through the carbon structure and the device operates as a perovskite LEC. In order to approve this assumption, the ultraviolet photoelectron spectroscopy (UPS) spectrums were measured SWCNT on top of perovskite with and without (reference) voltage exposure (figure S4 and table S1 in supplementary).

Carbon electrodes do not react with halides in perovskite components [25][40], which significantly enhances device life-time [41]. Single wall carbon nanotubes have properties such as low sheet resistant (<110 Ohm/sq) and high optical transparency (>90%) [42] and semiconductor behavior [43] [47]. SWCNTs are good electron/hole transport materials and electrode for perovskite optoelectronic devices [32,44]. SWCNT can be doped by halides (I, Br) and metals (Cs) with significant change of work function[36].

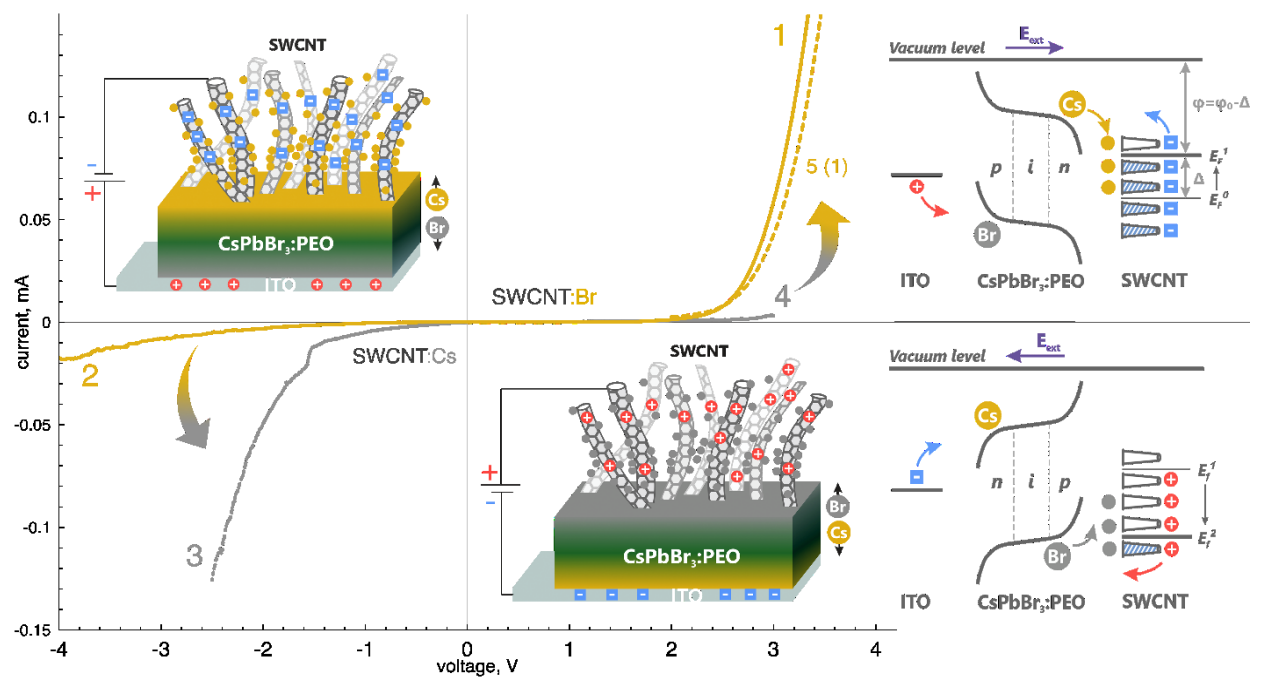

*Figure 1 — IV curve of single layer ambipolar perovskite LEC (ITO-CsPbBr3:PEO-SWCNT). Orange curves (1; 2 and 5) corresponds to forward sweep, SWCNT doped by Cs⁺ injects electrons; grey curves (3;4) correspond to switched mode: SWCNT doped by Br⁻ injects holes. Switching realize by the voltage induced ions migration. Energy diagram illustrate p-i-n and n-i-p formation and SWCNT doping in external electrical field: Br⁻ and Cs⁻ ions tune SWCNT by increasing and decreasing respectively work function and allow charge injection into perovskite.*

All device components including SWCNT are transparent in visible range (Figure 3 bottom). Transparent top electrode allows to fabricated stack tandem device with two multicolor subcells. Two red and green subpixels were linked face-to-face upon one with common SWCNT electrode. Glass-ITO substrates were covered with CsPbBr$_3$:PEO and CsPbI$_3$:PEO for green and red subpixels respectively. Stack configuration of subpixels takes less space than conventional side-by-side layout, which gain higher pixel density for high-resolution display application (Figure 2 — top: schematic illustration of conventional subpixel lauout against stack configuration and pixel architecture).

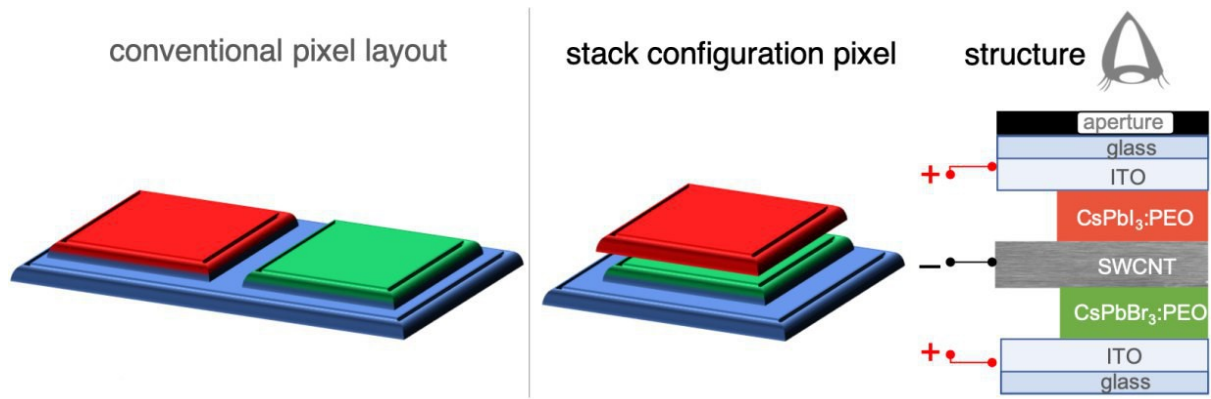

*Figure 2 — top: schematic illustration of conventional subpixel lauout against stack configuration and pixel architecture*

In stack configuration pixel both subcells operates independently. The variety of 522 nm and 691nm electroluminescence peaks demonstrated green-yellow-red emission colors (Figure 3). In this configuration the brightest cell (green) was shadowed by the red subcell as it shown on Figure 2. Fully assembled multicolor device is transparent in visible light range. Further optimization makes possible to increase whole transmittance and adopt this approach for transparent display technology for recently emerged AR devices.

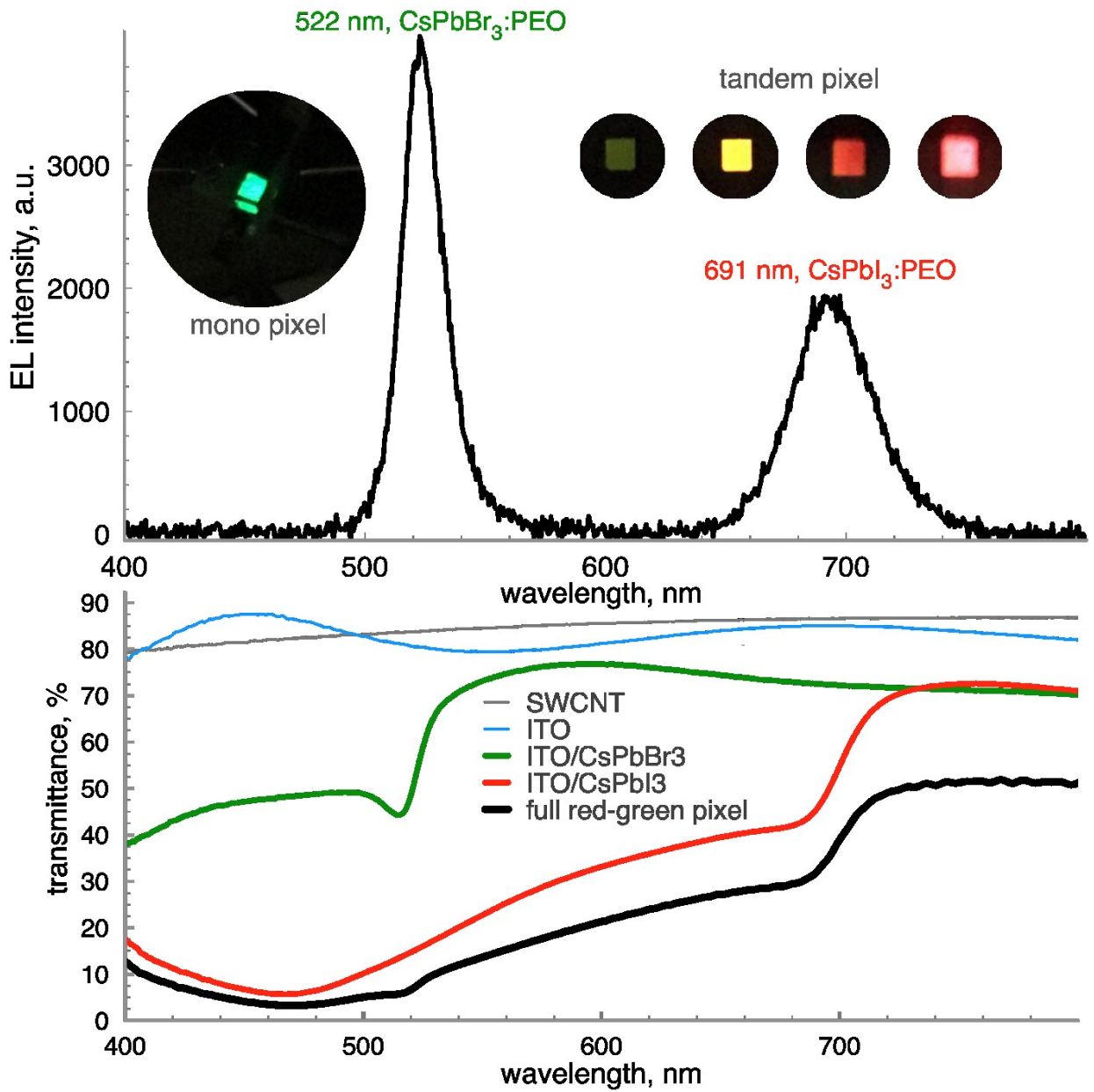

*Figure 3 — electroluminescence spectrum obtained by voltage, applied on red and green subcells in parallel connection. Colors variety obtained on stacked transparent subpixels. Bottom: transmittance spectrum of red-green stack pixel (black line) and each component of subpixels. and*

### Experimental section

The device architecture was inspired by the single layer perovskite LEC based on composite film of perovskite crystals in polyethylene oxide (PEO) [45] [23]. The device demonstrated eminent properties of maximum luminance (593 178 cd m$^{-2}$) and high performance (5.7% EQE) despite rough morphology of photoemissive layer of inorganic perovskite $CsPbBr_3$ and liquid indium-gallium electrode. Here we optimized the deposition process with the vacuum treatment for smooth uniform film. The vacuum treatment increases the evaporate rate of the solvents in the polymer-perovskite composite film, which quickly forms nucleation sites throughout the film before significant perovskite crystallization occurs.

Perovskite crystallization in films without vacuum treatment, proceed with a low nucleation site density. This leads to large crystals out growing far apart and results in nonuniform crystal domains with large gaps between perovskite crystals. Vacuum treatment of the wet films leads to the different way of crystallization to orthorhombic phase of CsPbBr$_3$ instead of cubic phase for non-vacuum crystallization. XRD pattern and SEM images are presented on Figure 1.

A light emissive film is composed of inorganic perovskite grains CsPbBr$_3$ (137.24 mg/ml synthesized by dissolving CsBr 63.84 mg/ml and PbBr$_2$ 73.4 mg/ml in dimethyl sulfoxide (DMSO)) in a polymeric polyelectrolyte matrix (PEO 10 mg/ml). All components dissolved individually in DMSO, mixed together overnight at 60°C. We observed that vacuum treatment of the wet film after spin coating significantly improves quality of perovskite film. Vacuum degassing quickly removes DMSO solvent from the polymer matrix, which solidifies the polymers and confine the perovskite precursors to separated domains. The separated domains impede the crystal growth and add nucleation centers to form a smooth and uniform film as compared to non-treated films. Improved morphology to prevents contact spots between the top and bottom electrodes and the device does not have leakage current. A pin-hole free film allows the use of planar electrodes instead of indium-gallium eutectic from [46].

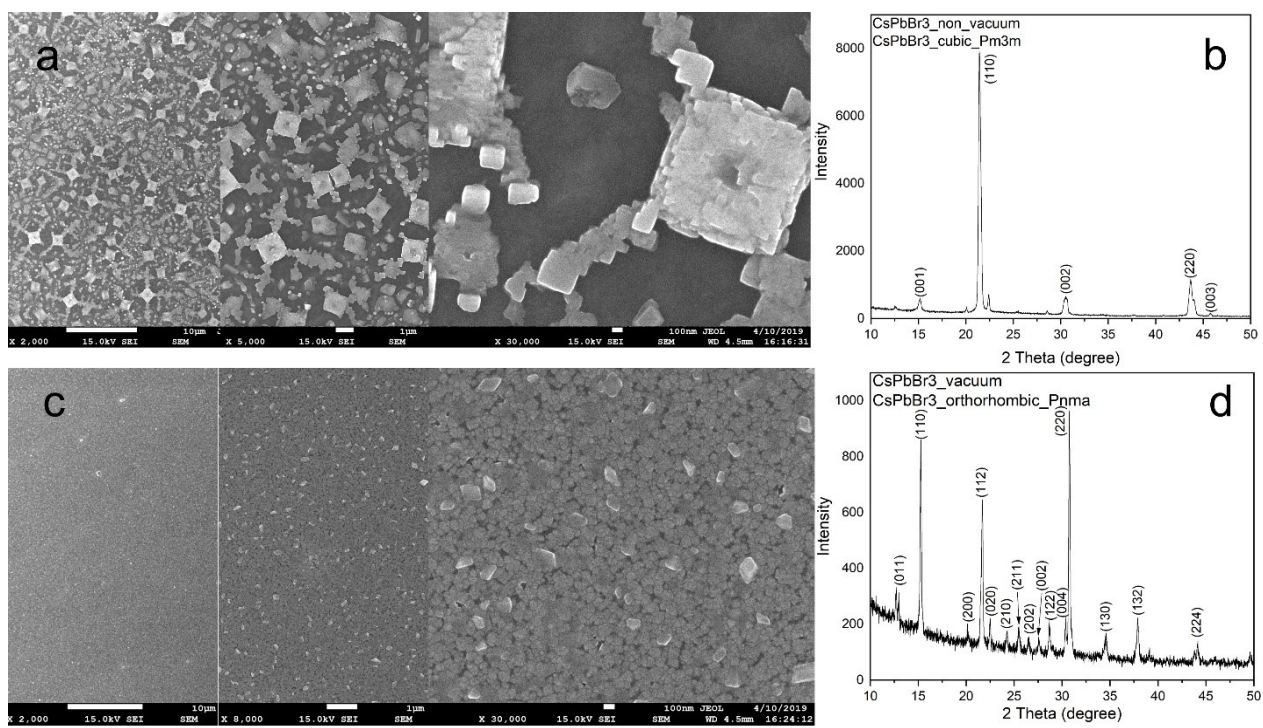

*Figure 4 — film characterization of CsPbBr3:PEO with vacuum treatment (c;b) and without (a, b) SEM imaging and XRD data*

Perovskite crystallization in films without vacuum treatment, proceed with a low nucleation site density. This leads to large crystals out growing far apart and results in nonuniform crystal domains with many pinholes. DMSO contained in wet polymer films, can evaporate immediately after contact with the hot surface of a heating plate after spin coating. The evaporated DMSO vapor leaves cavities in the solidified polymer. Different way of crystallization leads to orthorhombic phase of CsPbBr3 instead of cubic phase for non-vacuum crystallization. The comparison

Perovskite LECs were assembled using a glass substrate with ITO stripes as bottom electrode; spin-coated CsPbBr$_3$/I$_3$:PEO composite as emissive layer; SWCNT deposited by a simple press transfer process at room temperature as top electrode.

For this experiment high quality films of randomly oriented SWCNTs were produced by the aerosol (floating catalyst) chemical vapor deposition (CVD) method described in detail elsewhere [42]. SWCNT deposited by a simple press transfer process at room temperature as top electrode.

## Summary

In summary, we reported an ambipolar planar transparent perovskite LEC for the first time. Easy-to-do single layer inorganic perovskite light emitting electrochemical cell was optimized by vacuum treatment prior crystallization processed. Perovskite crystals imbedded polyelectrolyte matrix covered with transparent top carbon electrode deposited by dry lamination. Voltage induced reversable polarizing cycle tuned the device from p-i-n to n-i-p structure in-situ under electrical field. The combination of transparent, ionically doped carbon electrodes and polarization induced p-i-n formation allows electroluminescence in forward and reverse bias (ambipolar mode). Multicolor pixel was assembled in stack configuration based on semitransparent perovskite described LEC. The combination of green and red subcells perform yellow emission.

This concept demonstrates potential of perovskite LEC to advanced application in display devices. These features allow the possible fabrication of emissive multicolor micro-LED pixels in a stack configuration without a backlight. Stacked pixel designs have potential application in head-up displays (HUD) and augmented reality (AR) devices, due to smaller pixel area compared to conventional display with active matrix.

## Acknowledgements

The authors gratefully acknowledge the financial support of the Ministry of Science and Higher Education of the Russian Federation (№ К2-2019-014) and partial financial support Grant 14.Y26.31.0010.
.